# Examining the Effect of Monetary Policy and Monetary Policy Uncertainty on Cryptocurrencies Market

Mohammadreza Mahmoudi

*Department of Economics, Northern Illinois University, Dekalb, USA.*

Email: mmahmoudi@niu.edu

## ABSTRACT

This study investigates the influence of monetary policy and monetary policy uncertainties on Bitcoin returns, utilizing monthly data of BTC, and MPU from July 2010 to August 2023, and employing the Markov Switching Means VAR (MSM-VAR) method. The findings reveal that Bitcoin returns can be categorized into two distinct regimes: 1) regime 1 with low volatility, and 2) regime 2 with high volatility. In both regimes, an increase in MPU leads to a decline in Bitcoin returns: -0.028 in regime 1 and -0.44 in regime 2. This indicates that monetary policy uncertainty exerts a negative influence on Bitcoin returns during both downturns and upswings. Furthermore, the study explores Bitcoin's sensitivity to Federal Open Market Committee (FOMC) decisions.

## 1.  INTRODUCTION

Cryptocurrencies, emerging as an innovative financial asset over the last decade, have captured significant attention for their potential to revolutionize money, banking, and finance (Berman, 2023). The motivation to investigate the impact of monetary policy on this market arises from the noticeable interactions between established economic systems and the emerging digital economy. Specifically, after the economic difficulties following the COVID-19 pandemic, central banks globally have implemented new monetary policies, and the subsequent implications of these policies appear to have impacted the dynamics of the cryptocurrency markets (Cristina Polizu, 2023).

While there exists a substantial body of literature that has analyzed the effect of monetary policy on traditional stock markets, there is a noticeable gap in research concerning the impact of monetary policy on the cryptocurrency market. To fill this gap, the first part of the paper provides a broader picture to analyze the influence of Monetary Policy Uncertainty (MPU) on the cryptocurrency market dynamics using the Markov Switching Vector Autoregressive (MS-VAR) Model. This delves deeper into the subtle interactions, capturing potential regime shifts in the cryptocurrency market's response to varying degrees of MPU.

As depicted in Figure 1, the relationship between Bitcoin's return (BTC) and monetary policy uncertainty (MPU) become notably pronounced, particularly in the aftermath of the COVID-19 pandemic (Mahmoudi, 2023). Notably, the impact of recent monetary policies implemented to combat inflation post-COVID-19 has become inevitable in shaping Bitcoin's price and volatility. These observations underscore the extent to which cryptocurrencies, as integral components of the financial market, are closely intertwined with economic and monetary policies.

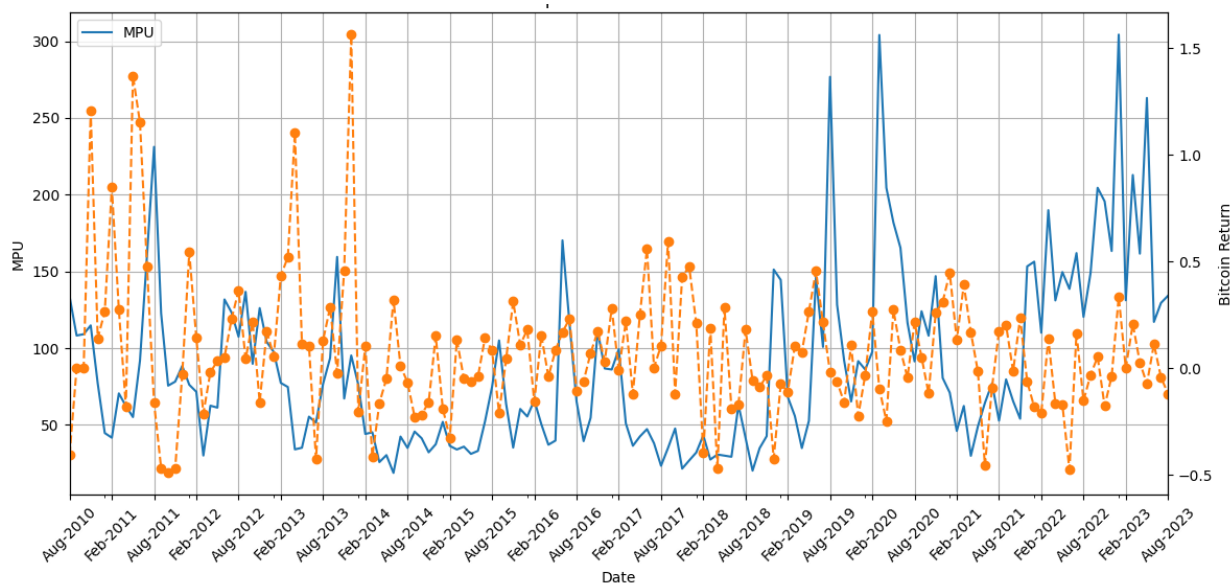

*Figure 1. Monthly Data of MPU and Bitcoin Return*
*Note: This plot represents monthly data of MPU and Bitcoin Return form July 2010 to August 2023.*

The relationship between monetary policy and its effects on the cryptocurrency market is not straightforward, particularly given the recent emergence of cryptocurrencies. A primary way this relationship works is through the interest rate channel. When institutions like the Federal Reserve change interest rates, they affect the cost of borrowing and the returns people get from saving (Mishkin, 1995). When interest rates are high, traditional investments such as bonds may offer better returns, reducing the appeal of cryptocurrencies. On the other hand, low rates might push investors to consider alternatives like cryptocurrencies.

Another channel is the wealth effect. When central banks use policies that increase asset prices, people who own these assets feel wealthier (Bernanke & Gertler, 1995). Feeling wealthier might make them more willing to invest in riskier assets, including cryptocurrencies. Similarly, when a country's currency becomes weaker due to certain monetary policies, cryptocurrencies might seem attractive as they can act as protection against this weakening currency (Eichenbaum

& Evans, 1995). Uncertainty around monetary policy can also influence decisions, with some viewing cryptocurrencies as a safe option during uncertain times (Bloom, 2009). Additionally, when there's more money available in the system because of certain policies, some of that money might find its way into cryptocurrencies. Central bank announcements, too, can affect perceptions about the economy's future (Romer & Romer, 2000). If these announcements make people concerned about the economy, they might consider holding more in cryptocurrencies. Finally, when the value of assets that can be used as collateral goes up, it can lead to increased borrowing. Some of this borrowed money might be invested in cryptocurrencies.

The structure of this paper unfolds as follows: Section 2 provides a comprehensive review of pertinent literature, shedding light on the existing research gaps. Section 3 presents the data. Section 4 outlines the research methodology, and statistical techniques. In Section 5, I present and analyze the empirical findings, while Section 6 concludes the paper by summarizing key insights and implications.

## 2. LITERATURE REVIEW

The relationship between monetary policy decisions and financial market responses has been a topic of persistent inquiry within the financial economics literature. While extensive research has been conducted on traditional assets like stocks, the burgeoning domain of cryptocurrencies remains less explored. This review aims to provide a concise overview of foundational literature on the stock market's reaction to Monetary Policy Uncertainty (MPU) and Monetary policy specifically Federal Open Market Committee (FOMC) announcements and subsequently highlight areas that merit further research, especially in the context of the cryptocurrency market.

### 2.1. Effect of Monetary Policy Uncertainties on Stock Markets

The Baker-Bloom-Davis MPU Indices serve as a systematic gauge for Monthly Monetary Policy Uncertainty (MPU) in the United States, using a methodological approach grounded in the analysis of newspaper articles (Baker et al., 2016). This analysis is predicated on the identification of specific keywords aligned with four designated criteria. The "E" criterion focuses on terms associated with the economic landscape, specifically "economic" and "economy." The "P" criterion encompasses terms tied to political and institutional realms, capturing words such as "congress," "legislation," "white house," "regulation," "federal reserve," and "deficit." The "U" criterion is centered on indicators of uncertainty, specifically the terms "uncertain" and "uncertainty." The "M" criterion, perhaps the most expansive, encapsulates terms directly related to monetary policy and its key figures. This includes "federal reserve," "the fed," "money supply," "open market operations," "quantitative easing," "monetary policy," "fed funds rate," "overnight lending rate," and names like "Bernanke," "Volker," and "Greenspan." It also comprises broader banking terms and institutions such as "central bank," "interest rates," "fed chairman," "fed chair," "lender of last resort," "discount window," "European Central Bank (ECB)," "Bank of England," "Bank of Japan (BOJ)," "Bank of China," "Bundesbank," "Bank of France," and "Bank of Italy." By systematically isolating articles that feature these keywords, the indices present a consistent and structured means to discern the levels of uncertainties associated with monetary policy for each month.

(Bekaert et al., 2013) investigated the stock market's response to unforeseen monetary policy changes. They determined that the stock market is indeed sensitive to these shifts. Their findings offer insights into how policy alterations can influence the financial market, notably in the context of market risks and movements. (Husted et al., 2020) delved deeper into Monetary Policy Uncertainty (MPU). They crafted a methodology to discern the distinct effects of MPU, with a particular emphasis on interest rates and the overarching financial market. Their contributions further illuminate our comprehension of MPU's impact on the economic landscape.

(Caggiano et al., 2020) broadened their scope to encompass both the U.S. and European financial markets. They sought to understand how policy adjustments in one region could resonate in another, underscoring the interconnected nature of global financial markets. In the realm of cryptocurrencies, the influence of MPU on their valuations is intricate. This influence is shaped by investor sentiment and various external determinants. (Kim, 2022) shows that the cryptocurrency market is connected to the traditional private money market through a type of cryptocurrency called reserve-backed stablecoins.

(Chowdhury et al., 2022) studied the effects of COVID-19 government interventions on the cryptocurrency market, particularly portfolio diversification. Using a Markov switching model, they found interventions negatively impacted cryptocurrency prices. The research recommends diversification strategies involving cryptocurrencies like Dogecoin, Stellar, and commodities like

Gold and Oil. They advise against heavy reliance on Bitcoin and Ethereum due to their strong correlation and encourage strategies to decrease market volatility. (Hsiao et al.) paper investigates the relationship between monetary policy uncertainty and cryptocurrency volatility. Surprisingly, the study reveals that, contrary to conventional wisdom, monetary policy uncertainty negatively predicts cryptocurrency volatility, with a more pronounced effect observed for longer time horizons, and it surpasses the performance of popular cryptocurrency volatility predictors.

In a notable study on Markov-switching VAR (MSVAR) models, (Kole & van Dijk, 2023) examined how economies, financial markets, and institutions navigate distress. They highlighted the MSVAR model's adeptness in integrating both gradual economic changes and abrupt regime shifts. This approach is particularly useful for analyzing the ramifications of shocks during recessions or bearish financial periods. Introducing a methodology that captures primary and secondary statistical moments based solely on regime distribution, they further proposed impulse response functions for these moments. By adapting the MSVAR into an extended linear non-Gaussian VAR format, they ensured concise and interpretable results. Their research, when applied to stock and bond return predictability, revealed that forecasts, such as means, volatilities, and autocorrelations, vary based on the prevailing regime. Notably, during bear markets, shocks exerted amplified and more prolonged effects on means and volatilities.

Existing literature, as exemplified by works from (Baker et al., 2019); (Raunig, 2021) ; (Chen et al., 2023) ; (Ajmi et al., 2015) ; (Batabyal & Killins, 2021)), predominantly explores the ramifications of monetary policy uncertainties within the realm of conventional financial markets.

While extensive research has outlined MPU's influence on traditional markets, a noticeable void exists regarding its effects on emerging platforms like cryptocurrencies. As digital currencies rise in significance, understanding their interplay with core economic factors, such as MPU, becomes essential. Although much of the existing research is centered on the U.S. and Europe, a holistic, global examination is necessary. Furthermore, the introduction of cutting-edge research methodologies opens up fresh perspectives on this topic. Accordingly, in the initial segment of this study, I analyze the impact of MPU on the cryptocurrency market using the Markov regime-switching VAR model.

## 2.2. FOMC Announcements and Stock Market Responses

In a more detailed analysis, (Bomfim, 2003) by considering both the macroeconomic insights embedded in FOMC announcements and the actual monetary policy decisions. This dual approach enhances our understanding by differentiating the impact of news and policy changes on stock market movements. Broadening the scope, (Rigobon & Sack, 2004) emphasized the profound influence of monetary policy on equities. Their work aids a holistic appreciation of how asset markets interpret and adapt to monetary policy cues.

(Bernanke & Kuttner, 2005) pioneered the study of equity price reactions to unexpected alterations in the federal funds rate. Their results demonstrate that unexpected shifts in this rate directly affect stock valuations, establishing a fundamental link between monetary policy choices and stock market outcomes. (Lucca & Moench, 2015) delved into the proactive behaviors in the stock market, pointing out a noteworthy trend in U.S. stock prices during the 24-hour window preceding FOMC announcements. This insight accentuates the market's inclination to forecast and adjust to anticipated monetary policy changes.

(Che et al., 2023) investigation centers on the oscillations in the cryptocurrency market and their associations with global equity markets and U.S. monetary policy. The study recognizes a primary "crypto factor" accounting for 80% of cryptocurrency price shifts and spotlights its rising association with equity markets after institutional investors ventured into the cryptocurrency realm. Notably, the findings show that U.S. Federal Reserve tightening diminishes the crypto factor, challenging the belief that cryptocurrencies serve as a safeguard against market uncertainties.

Despite extensive studies on traditional assets in relation to FOMC announcements, the realm of cryptocurrencies remains largely uncharted. Given the distinctive and often turbulent nature of cryptocurrencies, focused research is essential.

- Differential Asset Behavior: Cryptocurrencies have inherent features distinct from traditional stocks. Their speculative essence, together with decentralized structures, demands a renewed investigative approach.

- Market Maturity and Global Influence: Cryptocurrencies, being universally traded with a relatively nascent market presence, may exhibit unique reactions to predominantly U.S.-oriented FOMC announcements. This global facet, paired with market novelty, highlights potential discrepancies in market responses.

This paper sets itself apart by utilizing a Markov regime-switching methodology to assess the influence of MPU on Bitcoin, shedding light on their intricate interrelation. Moreover, I explore the repercussions of FOMC resolutions on Bitcoin's returns and volatility, presenting an in-depth assessment of monetary policy's bearing on the cryptocurrency sphere. Such distinct features of my research enhance our grasp of cryptocurrency market intricacies.

## 3. DATA

In this research, the initial section involves the utilization of monthly data encompassing BTC and MPU from July 2010 to August 2023. I sourced the Bitcoin price data from (CoinDesk, 2023) and obtained the monetary policy uncertainty (MPU) data from the (MPU, 2023). Table 1 provides key data statistics for monetary policy uncertainty (MPU) and Bitcoin price (BTC) in the level and natural log returns forms. BTC exhibits notable variability, with a mean and standard deviation of 9,324.254 and 14,5558.63, respectively. The highest recorded index reached 60,955.77 in November 2021, while the lowest was 0.06 in August 2010. In contrast, MPU displays lowest volatility with respect to other indexes, featuring a mean and standard deviation of 88.82547 and 57.56636, respectively. The lowest figure observed was 18.68333 in August 7, 2014, whereas the highest was 304.2905 in May 2020.

*Table 1. Descriptive Statistics of MPU, and BTC in Level and Return Form*

| | Mean | Median | Standard Deviation | Kurtosis | Skewness | Range | Minimum | Maximum | Count |
|---|---|---|---|---|---|---|---|---|---|
| **Panel A: Level Data** | | | | | | | | | |
| **BTC** | 9324.254 | 992.52 | 14558.63 | 2.897295 | 1.896145 | 60955.71 | 0.06 | 60955.77 | 158 |
| **MPU** | 88.82547 | 74.73636 | 57.56636 | 2.147595 | 1.372203 | 285.6071 | 18.68333 | 304.2905 | 158 |
| **Panel B: Natural Log Returns** | | | | | | | | | |
| **BTC** | 0.080071 | 0.043972 | 0.324766 | 4.936772 | 1.584392 | 2.052308 | -0.48993 | 1.562375 | 157 |
| **MPU** | -0.0011 | -0.04415 | 0.449795 | 0.340706 | 0.447324 | 2.323059 | -0.86853 | 1.454527 | 157 |

*Note: Panel A details the descriptive statistics for levels of MPU and BTC. Panel B describes the statistics for the returns of these entities. BTC shows significant fluctuations with a mean of 9,324.254 and a deviation of 14,5558.63. Conversely, MPU has the least volatility among the indices, with a mean of 88.82 and a deviation of 57.56.*

Figure 2. Level and return quantities of BTC and MPU illustrates a graph showcasing the level values of MPU and Bitcoin price in the left side, as well as the return/growth quantities of BTC and MPU in the right side. Notably, our examination is based on the computed return, defined by the formula: $return = ln(\frac{P_t}{P_{t-1}})$.

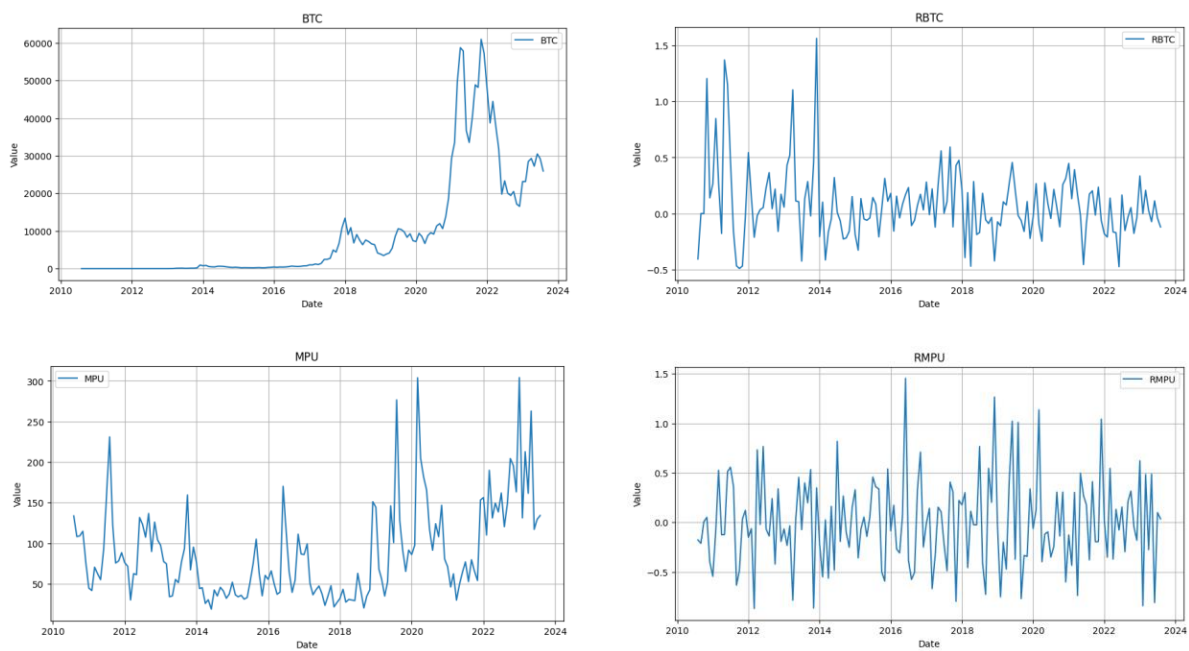

*Figure 2. Level and return quantities of BTC and MPU*

*Note: The left panels show the plot of natural logarithms of MPU and Bitcoin price as well as the right panels depict the returns of BTC and MPU.*

Towards the end of my paper, I explore the effects of the Federal Open Market Committee (FOMC) decisions on Bitcoin returns. I sourced FOMC data from the (Reserve, 2023) and used hourly Bitcoin prices from (CoinDesk, 2023). Table 2 provides a summary of key statistics related to FOMC interest rate changes and hourly Bitcoin returns. The average change in the interest rate by the FOMC is around 0.1 percent with a standard deviation of 0.3 percent. The most significant decrease in the interest rate was 1 percent on 2020-01-29 at 15:00:00. In contrast, the largest increases were 0.75 percent, noted on dates including 2022-06-15, 2022-07-27, 2022-09-21, and 2022-11-02, all at 14:00:00. As for Bitcoin, the average return is 3.95E-05 with a standard deviation of 0.7 percent. Remarkably, Bitcoin's highest hourly return was 18 percent, and its lowest was -16 percent.

*Table 2. Descriptive Statistics of hourly data of Bitcoin returns and interest rate change based on FOMC decision*

| | Mean | Median | Standard Deviation | Kurtosis | Skewness | Range | Minimum | Maximum | Count |
|---|---|---|---|---|---|---|---|---|---|
| **FOMC** | 0.000864 | 0 | 0.002895 | 3.87536 | -0.20209 | 0.0175 | -0.01 | 0.0075 | 55 |
| **BTC** | 0.060184 | 0.042099 | 0.31164 | 0.065426 | 0.222969 | 1.45024 | -0.56456 | 0.885677 | 55 |

*Note: This table provides a summary of key statistics from hourly data on Bitcoin returns and interest rate changes as decided by the FOMC. The FOMC's average interest rate change stands at 0.1 percent, with a variability of 0.3 percent. In comparison, Bitcoin's mean return is 3.95E-05, accompanied by a standard deviation of 0.7 percent.*

## 4. METHODOLOGY

To study the effect of MPU on Bitcoin returns, we need models that can handle the complex links between our variables. Simple models might not capture these fully. Because of this and the nature of my data, I will use Markov Switching Vector Autoregressive models (MS-VAR). The Markov Switching Vector Autoregressive (MS-VAR) model is particularly suited for my analysis on the impacts of MPU on Bitcoin returns. This model stands out as it can capture the intricate relationships and interactions between these multiple variables, understanding their interconnected dynamics over time. Unlike some other models, the MS-VAR allows for shifts in these relationships, which is essential given the ever-changing nature of financial markets. It can pinpoint times of heightened volatility or changing influences between the policy uncertainties and Bitcoin. Moreover, the model's ability to consider possible regime changes both in terms of mean and variance gives it an edge, offering a more comprehensive view of the policy uncertainties' impact on Bitcoin returns in different market conditions.

The general MS-VAR model of order q with r regime for a time series variable $x_t$ is illustrated below (Krolzig, 1998):

$$x_t = \begin{cases} \alpha_1 + \beta_{11}x_{t-1} + \ldots + \beta_{q1}x_{t-q} + \sigma_1^{1/2}u_t, & if\ s_t = 1 \\ \alpha_m + \beta_{1m}x_{t-1} + \ldots + \beta_{qr}x_{t-q} + \sigma_r^{1/2}u_t, & if\ s_t = r \end{cases}$$

$$u_t | s_t \sim NID(0, I_k) \qquad\qquad (1)$$

This equation can be expressed as follows:

$$x_t = \alpha_{s_t} + \sum_{i=1}^{q} \beta_{i,s_t}x_{t-i} + \sigma_{s_t}^{1/2}u_t \qquad\qquad (2)$$

Where $x_t = \begin{pmatrix} x_1 \\ \vdots \\ x_p \end{pmatrix}, \beta_i = \begin{pmatrix} \beta_{11} & \cdots & \beta_{1q} \\ \vdots & \ddots & \vdots \\ \beta_{q1} & \cdots & \beta_{qq} \end{pmatrix}, x_{t-i} = \begin{pmatrix} x_{t-1} \\ \vdots \\ x_{t-p} \end{pmatrix}, \alpha_{s_t} = \begin{pmatrix} \alpha_{1,s_t} \\ \vdots \\ \alpha_{1.s_t} \end{pmatrix}$

$\beta_i$ represents the matrix of autoregressive coefficients. $\alpha$ is the intercept term, $\sigma^{1/2}$ denotes the standard deviations dependent on $s_t$, the regime at time t.

I assume the error term, $u_t$, in equations 1 and 2 follows a normal distribution, leading to the conditional probability density function for $x_t$ as follows:

$$p(x_t|s_t = t_m, X_{t-1}) = ln\left(2\pi^{-\frac{1}{2}}\right) ln\left|\sigma^{-\frac{1}{2}}\right| exp\left\{(x_t - \overline{x_{rt}})'^{\sigma_m^{-1}}(x_t - \overline{x_{rt}})\right\} \qquad (3)$$

where $\overline{x_{mt}} = E[x_t|s_t, x_{t-1}]$, and $X_{t-1} = (x'_{t-1}, x'_{t-2}, \ldots, x'_1, x'_0, \ldots, x'_{1-q})'$

Then, the conditional density function of $x_t$ is normal, $x_t|s_t = t_r, X_{t-1} \sim NID(\overline{x_{rt}}, \sigma_r)$

$$\overline{x_t} = \begin{bmatrix} \overline{x_{1t}} \\ \vdots \\ \overline{x_{rt}} \end{bmatrix} = \begin{bmatrix} \alpha_1 + \sum_{i=1}^{q} \beta_{1,i} x_{t-i} \\ \vdots \\ \alpha_r + \sum_{i=1}^{q} \beta_{1,r} x_{t-i} \end{bmatrix} \qquad (4)$$

Underlying these equations is the assumption that the state variables, $s_t$, which determine the switching behavior of the time series variable $x_t$, follow an irreducible ergodic two-state Markov process. This assumption implies that the current regime $s_t$ depends on the regime one period earlier, $s_{t-1}$. Consequently, the transition probability between states is given by:

$$P(S_t = i|S_{t-1} = j, S_{t-2} = k, \ldots) = P(S_t = i|S_{t-1} = j) = p_{ji} \qquad (5)$$

Where $p_{ji}$ represents the transition probability from state j to state i. Typically, this transition probability is represented by an ($n \times n$) matrix as follows:

$$P = \begin{bmatrix} p_{11} & p_{12} & \cdots & p_{1r} \\ p_{21} & p_{22} & \cdots & p_{2r} \\ \vdots & \vdots & \ddots & \vdots \\ p_{r1} & p_{r2} & \cdots & p_{rr} \end{bmatrix} \tag{6}$$

$$\sum_{j=1}^{r} p_{ij} = 1, i = 1,2,\ldots,r, \text{and } 0 \leq P_{ij} \leq 1$$

Various Markov-switching vector autoregression models can be employed to analyze the influence of monetary policy uncertainties on cryptocurrency market, as summarized in **Error! Reference source not found.**. In essence, there are two primary types of Markov-switching VAR models: Markov-switching mean (MSM) and Markov-switching intercept (MSI). It is important to note that MSM and MSI models differ in their behavior. A rapid shift in the observed time series vector to a new level is driven by the dynamic response to a regime change in the mean, denoted as $\mu(s_t)$. In contrast, a regime shift in the intercept term, $\alpha(s_t)$, results in a dynamic response similar to that of a comparable shock in the white noise series $u_t$ (Mahmoudi, 2022).

*Table 3. Different Markov-switching VAR models*

| | | MSI | | MSM | |
|---|---|---|---|---|---|
| | | **α invariant** | **α varying** | **μ invariant** | **μ varying** |
| **$\beta_j$ Invariant** | **$\sigma$ invariant** | MSI-VAR | Linear VAR | MSM-VAR | Linear MVAR |
| | **$\sigma$ varying** | MSIH-VAR | MHA-VAR | MSMH-VAR | MSH-MVAR |
| **$\beta_j$ Varying** | **$\sigma$ invariant** | MSIA-VAR | MSA-VAR | MSMA-VAR | MSA-MVAR |
| | **$\sigma$ varying** | MSIAH-VAR | MSAH-VAR | MSMAH-VAR | MSAH-MVAR |

*Note: This table contains all Markov-switching Vector Autoregressive Models. M indicates Markov-switching mean, I represents Markov-switching intercept, A shows Markov switching autoregressive, and H determines Markov-switching heteroskedasticity* (Krolzig, 1998)

## 5.  EMPIRICAL RESULTS AND DISCUSSION

I conducted stationarity tests employing the Augmented Dickey Fuller (ADF), Phillips–Perron, and Kwiatkowski-Philips-Schmidt-Shin (KPSS) methods. According to the findings presented in Table 4, it is observed that MPU exhibits non-stationary behavior when analyzed in all forms, whereas Bitcoin only demonstrate stationarity when considered in terms of returns.

*Table 4. Unit root test for levels, natural logarithms and returns of MPU and BTC*

|  |  | **ADF** | **Phillips–Perron** | **KPSS** |
|---|---|---|---|---|
| **Level** | **BTC** | -1.479594 (0.5414) | -1.377783 (0.5919) | 0.992190 |
|  | **MPU** | -4.098940 (0.0013) | -5.927832 (0.0000) | 0.542429 |
| **Natural Log** | **LBTC** | -3.258061 (0.0186) | -3.287511 (0.0171) | 1.41156 |
|  | **LMPU** | -4.786227 (0.0001) | -4.670361 (0.0002) | 0.439840 |
| **Returns/ Growth** | **RBTC** | -9.812166 (0.0000) | -9.807175 (0.0000) | 0.41063 |
|  | **RMPU** | -15.41783 (0.0000) | -27.16840 (0.0000) | 0.28058 |

*Note: This table presents tests (ADF, Phillips–Perron, and KPSS) to check the stationarity of values, logs, and returns of MPU and Bitcoin. The findings indicate that Bitcoin's values and logs are not stationary, but its returns are. On the other hand, MPU is stationary in all ways. The numbers in brackets show the p-value.*

Subsequently, our aim was to identify a lasting relationship among BTC return and MPU. To assess cointegration within our dataset, the first step was to determine the optimal lag. To achieve this, I conducted a VAR model analysis. The Vector Autoregression (VAR) model is a stochastic process model employed to comprehend the linear relationships among various time series data. It expands upon the univariate autoregressive model (AR model) by incorporating multiple evolving variables. An example of the VAR(p) Model is as follows:

$$X_t = \alpha + \beta_1 X_{t-1} + \beta_2 X_{t-2} + \ldots + \beta_q X_{t-q} + \varepsilon_t$$

As evident from the information presented in Table 5, the optimal lag duration, determined by considering the AIC criteria, is 1.

*Table 5. VAR lag order selection criteria*

| Lag | LogL | FPE | AIC | SC | HQ |
|-----|------|-----|-----|-----|-----|
| 0 | -130.4391096769548 | 0.020281 | 1.777706 | 1.818028 | 1.794088 |
| 1 | -119.2906569115331 | 0.018426* | 1.681754* | 1.802718* | 1.730900* |
| 2 | -118.6387455870841 | 0.019274 | 1.726695 | 1.928302 | 1.808604 |
| 3 | -114.3541699773635 | 0.019202 | 1.722875 | 2.005125 | 1.837548 |
| 4 | -110.3280414708298 | 0.019198 | 1.722524 | 2.085417 | 1.869961 |
| 5 | -107.4064004365713 | 0.019483 | 1.736999 | 2.180534 | 1.917200 |
| 6 | -100.9208375014004 | 0.018850 | 1.703635 | 2.227814 | 1.916600 |
| 7 | -100.5250803182329 | 0.019794 | 1.752015 | 2.356836 | 1.997743 |
| 8 | -95.30210506109879 | 0.019484 | 1.735599 | 2.421063 | 2.014092 |

*Note: This table summarized several criteria to determine the optimal lag length of VAR model. Based on different criteria, the optimal lag for our data is 2.*

After identifying the optimal lag, the subsequent step involves conducting the cointegration test. The Johansen test is employed for assessing cointegration and enables the evaluation of multiple cointegrating relationships. The Johansen test estimates the rank (r) of a given time series matrix while considering a specified confidence level. As indicated in Table 6, the Johansen test evaluates two null hypotheses: one where r=0, indicating no cointegration, and another where r≤1, signifying the presence of at most one cointegration relationship. For the case of r=0, all test values fall below the critical values, leading to the inability to reject the null hypothesis, implying the absence of any cointegration relationship. However, in the case of r≤1, none of the test values exceed the critical values, leading to the rejection of the null hypothesis. Thus, there is no cointegration relationship among our variables.

*Table 6. Johansen Cointegration Test*

|  | Test Statistics | Critical Values | | |
|---|---|---|---|---|
|  |  | 10 percent | 5 percent | 1 percent |
| **Panel A: Trace Test** | | | | |
| **r = 0** | 11.98 | 15.66 | 17.95 | 23.52 |
| **r ≤ 1** | 0.86 | 6.5 | 8.18 | 11.65 |
| **Panel B: Maximum Eigenvalue** | | | | |
| **r = 0** | 11.12 | 12.91 | 14.9 | 19.19 |
| **r ≤ 1** | 0.86 | 6.5 | 8.18 | 11.65 |

*Note: This table summarized the Trace and Maximum Eigenvalue tests in order to evaluate cointegration relationships in our data. The results of different tests show there is no cointegration relations between Bitcoin price and MPU.*

In the absence of cointegration, the dependent variables deviate from any conceivable combination of the explanatory variables. Consequently, conducting a Vector Autoregression (VAR) analysis in its original levels is not advisable (Juselius, 2006). When one or all of the variables in a regression exhibit integrated properties (I(1)), standard statistical results may or may not hold. The concept of spurious regression serves as a notable example, where the typical statistical results do not apply, even when all the regressors are I(1) and lack cointegration (Zivot & Wang, 2006). To avoid encountering spurious regression and to explore potential nonlinearities in the relationship between monetary policy uncertainties and Bitcoin returns, I employ a Markov-switching VAR model. According to the AIC criteria, the most suitable model that aligns with our data is MSM(2)-VAR(1) model. Markov Switching Means VAR (MSM-VAR) model is an extension of the Vector Autoregressive (VAR) model, allowing for regime-switching capabilities in the intercepts or "means" of the model while keeping the dynamics (i.e., the coefficients on the lags of the variables) unchanged across regimes. In essence, this model captures the potential changes in the means or levels of the series under study, based on different states or regimes, which are determined by an underlying Markov process (Mahmoudi & Ghaneei, 2022).

I used the MSM(2)-VAR(1) model to study the Bitcoin returns. Based on Table 7 information, Bitcoin return has two regimes: a high volatility (regime 1), and a low volatility (regime 2).

*Table 7. MSM(2)-VAR(1) Model of Bitcoin Return*

|  |  | $RBTC_t$ |
|---|---|---|
| **Intercepts** | **Const (Regime 1)** | 0.027 (0.282) |
|  | **Const (Regime 2)** | 0.286 (0.08) |
| **Volatilities** | **Sigma (Regime 1)** | 0.242 (0.011) |
|  | **Sigma (Regime 2)** | 0.513 (0.0002) |
| **Autoregressive Coefficients** | **AR(1)** | 0.127 (0.255) |
|  | **AR(2)** | 0.05 (0.53) |

*Note: This table summarized the results of Bitcoin return's structure using MSM(2)-VAR(1) Model. $RBTC_t$ implies natural logarithms of Bitcoin return at time t. The results show Bitcoin return has two regimes including low volatility (regime 1) and high volatility (regime 2). The values in parenthesis represent p-value.*
*Significance. codes: 0 '***' 0.001 '**' 0.01 '*' 0.05 '.' 0.1 ' ' 1*

I calculated the transition probabilities of the MSM(2)-VAR(1) Model concerning Bitcoin returns, as illustrated in Table 8. The value of $P_{11}$, denoting the likelihood of remaining in regime 1 when already in regime 1, stands at 89%. Meanwhile, $P_{21}$, symbolizing the transition likelihood from regime 2 to regime 1, is 3%.

*Table 8. Transition Probabilities of MSI (2)-VAR (1) Model for Bitcoin Return*

|  | **Regime 1** | **Regime 2** |
|---|---|---|
| **Regime 1** | 0.895 | 0.105 |
| **Regime 2** | 0.03 | 0.97 |

*Note: This table summarized the transition probabilities of Bitcoin return using MSM(2)-VAR (1). For example, $P_{11}$, the probability of staying in regime 1 given that you are in regime 1, is 89%, and $P_{21}$, the coefficients for the transition from regime 2 into regime 1, is 3%.*

The filtered probabilities of being in two regimes for Bitcoin return is depicted in Figure 3.

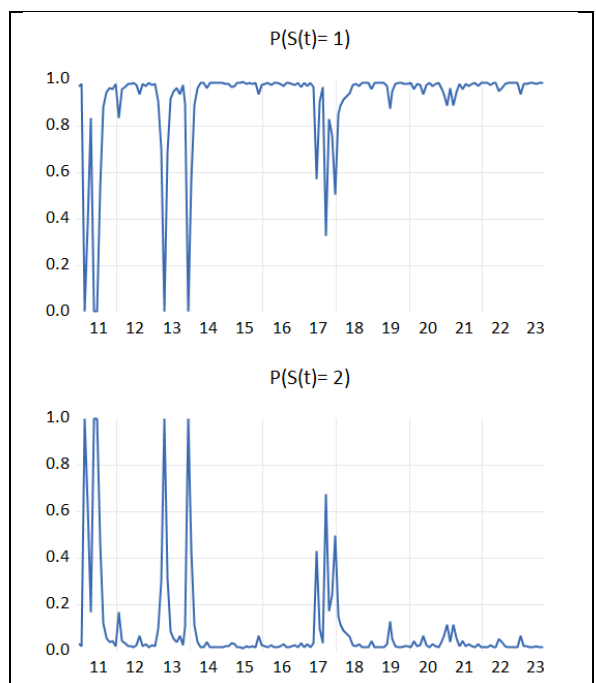

*Figure 3. Markov switching Filtered Regime Probabilities of Bitcoin return*

Note*: The graph illustrates the Markov-switching filtered probabilities for Bitcoin's returns across two regimes: regime 1, characterized by low volatility, and regime 2, marked by high volatility.*

Based on the AIC criteria, the MSM(2)-VAR(1) Model is optimal for examining how monetary policy uncertainties (MPU) influence Bitcoin returns. Table 9 demonstrates that in regime 1, which has low volatility, the average monthly Bitcoin return is 0.024. During this regime, monthly changes for MPU is 0.018. In contrast, regime 2 sees a monthly Bitcoin return of 0.336, with MPU changing by 0.062. Regime 1, or the bull market, is less volatile than regime 2, the bear market. Interestingly, in both regimes an uptick in MPU reduces Bitcoin returns by -0.028 and -0.44 in regime 1 and regime 2 respectively. This suggests monetary policy uncertainty has a negative effect during bearish and bullish periods.

Table 9. The effect of policy uncertainties on Bitcoin return using MSM (2)-VAR (1) Model (Based on AIC criteria)

| | | $RBTC_t$ | $RMPU_t$ |
|---|---|---|---|
| **Intercept** | **Const (Regime 1)** | 0.024 (0.247) | 0.018 (0.599) |
| | **Const (Regime 2)** | 0.336 (0.046) | 0.062 (0.403) |
| **Standard Deviation** | **Sigma (Regime 1)** | 0.04 (<0.0001) | 0.198 (<0.0001) |
| | **Sigma (Regime 2)** | 0.237 (0.0004) | 0.129 (0.0008) |
| **Regime 1** | $RBTC_{t-1}$ | 0.0538 (0.604) | 0.083 (0.65) |
| | $RMPU_{t-1}$ | -0.028 (0.492) | -0.24 (0.006) |
| **Regime 2** | $RBTC_{t-1}$ | 0.381 (0.187) | 0.358 (0.068) |
| | $RMPU_{t-1}$ | -0.44 (0.143) | -0.16 (0.381) |

*Note: This table summarized the effect of monetary policy uncertainties (MPU) on Bitcoin return using MSM(2)-VAR(1). $RBTC_t$ and $RMPU_t$ indicate natural logarithms of return of Bitcoin, MPU at time t. The results reveal that in regime 1, Bitcoin's monthly return averages 0.024 with varying changes in MPU. In the more volatile regime 2, Bitcoin's return increases by 0.336 monthly, with policy uncertainties affecting returns differently across bull and bear markets. In both regimes an uptick in MPU reduces Bitcoin returns by -0.028 and -0.44 in regime 1 and regime 2 respectively. This suggests monetary policy uncertainty has a negative effect during bearish and bullish periods. Significance. codes: 0 '***' 0.001 '**' 0.01 '*' 0.05 '.' 0.1 ' ' 1*

Furthermore, the transition probabilities of MSM(2)-VAR(1) Model for the effect of monetary policy uncertainties on Bitcoin return is shown in Table 10. Based on this information, when we consider the effect of policy uncertainties, $P_{11}$, which represents the probability of staying in regime 1 given that you are in regime 1, is 71 percent, as well as $P_{21}$, which indicates the coefficients for the transition from regime 1 into regime 2, is 8 percent.

*Table 10. Transition Probabilities of MSM(2)-VAR(1) Model for the effect of MPU on Bitcoin return*

| | Regime 1 | Regime 2 |
|---|---|---|
| **Regime 1** | 0.7167 | 0.2832 |
| **Regime 2** | 0.08267 | 0.917 |

*Note: This table displays the transition probabilities of the MSM(2)-VAR(1) Model concerning the influence of policy uncertainties on Bitcoin return. Notably, P_11, the likelihood of remaining in regime 1 when already in it, stands at 71%. Meanwhile, P_21, the probability of transitioning from regime 1 to regime 2, is 8%.*

Additionally, the response of BTC to MPU in both regimes is illustrated in Figure 4. As evident from the plots of both regimes, BTC's initial response to MPU is negative in both cases, after which it reverts to its long-term trend.

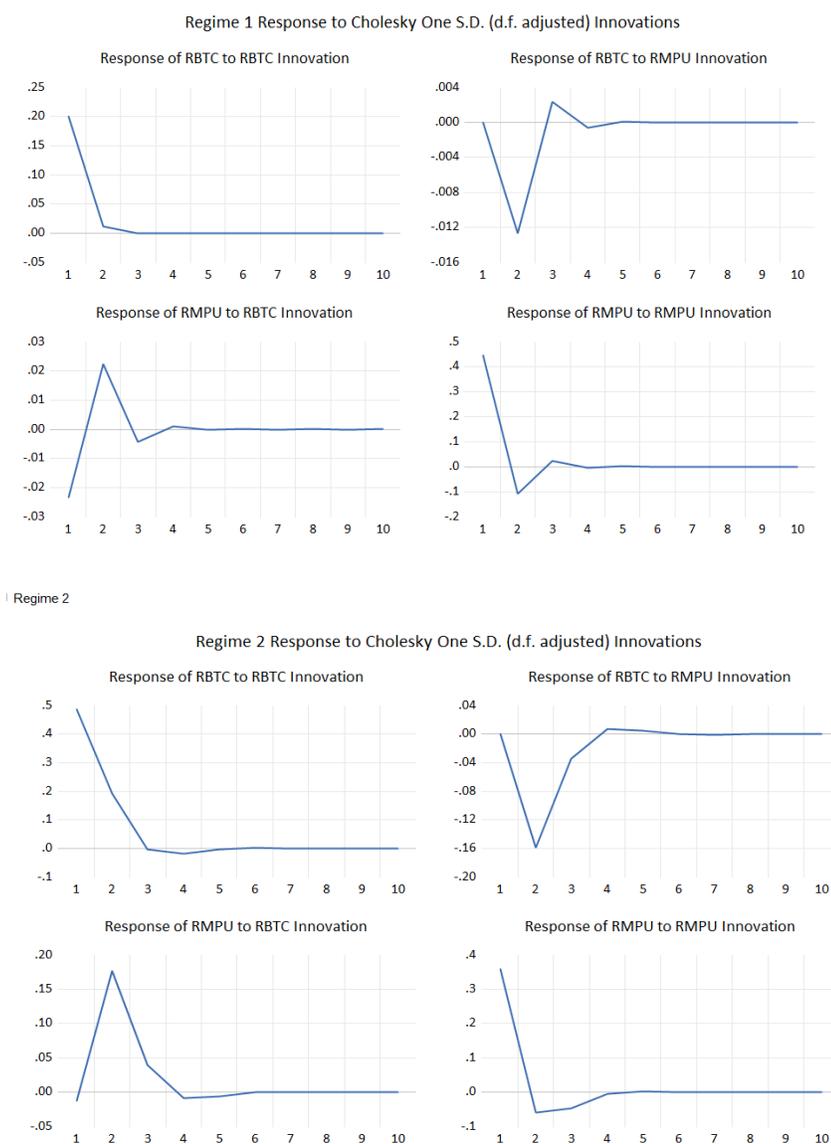

*Figure 4. Impulse responses of BTC to MPU innovative in both regime 1 and regime 2*

*Note: This figure represents the impulse responses of BTC to MPU in both regime 1 and regime 2. As it clear, in both regimes, Bitcoin return responded negatively to MPU innovations.*

## 5.1. ANALYZE THE IMPACT OF THE MONETARY POLICY ON
## CRYPTOCURRENCIES MARKET

In this section, I will explore how monetary policy influences the cryptocurrency market. Specifically, I'll examine the effect of the Federal Open Market Committee (FOMC) decisions on Bitcoin returns and volatility. For this analysis, I'll use hourly Bitcoin price data spanning from January 2017 to August 2023. The Federal Open Market Committee (FOMC) is the Federal Reserve's monetary policy-making body. The Federal Open Market Committee (FOMC) is the branch of the Federal Reserve responsible for shaping monetary policy. They manage the nation's open market activities, including the purchase and sale of U.S. Treasury securities. Through their actions, the FOMC affects the fed funds rate, which in turn influences other interest rates in the economy. The last 10 rows of FOMC dataset is shown at Table 11.

*Table 11. The FOMC Data*

|    | Datetime | Actual | Forecast | Previous |
|----|----------|--------|----------|----------|
| 45 | 2022-07-27 14:00:00 | 0.0250 | 0.0250 | 0.0175 |
| 46 | 2022-09-21 14:00:00 | 0.0325 | 0.0325 | 0.0250 |
| 47 | 2022-11-02 14:00:00 | 0.0400 | 0.0400 | 0.0325 |
| 48 | 2022-12-14 15:00:00 | 0.0450 | 0.0450 | 0.0400 |
| 49 | 2023-02-01 15:00:00 | 0.0475 | 0.0475 | 0.0450 |
| 50 | 2023-03-22 14:00:00 | 0.0500 | 0.0500 | 0.0475 |
| 51 | 2023-05-03 14:00:00 | 0.0525 | 0.0525 | 0.0500 |
| 52 | 2023-06-14 14:00:00 | 0.0525 | 0.0525 | 0.0525 |
| 53 | 2023-07-26 14:00:00 | 0.0550 | 0.0550 | 0.0525 |
| 54 | 2023-09-20 14:00:00 | 0.0550 | 0.0550 | 0.0550 |

*Note: this Table shows the Last 10 rows of FOMC data which contains the Date and time of FOMC announcement, Actual interest rate at that date, the forecast for the time, and the Previous interest rate.*

During FOMC meeting days, announcements are made regarding whether the federal funds rate has increased, decreased, or remained unchanged. The bar plot of Actual, Previous, and Forecast FOMC interest rate is depicted in Figure 5. Over the given timeframe, the FOMC convened 55 times. In these gatherings, they increased the interest rate 18 times, decreased it 5 times, and opted to maintain it at the current level 32 times (Reserve, 2023).

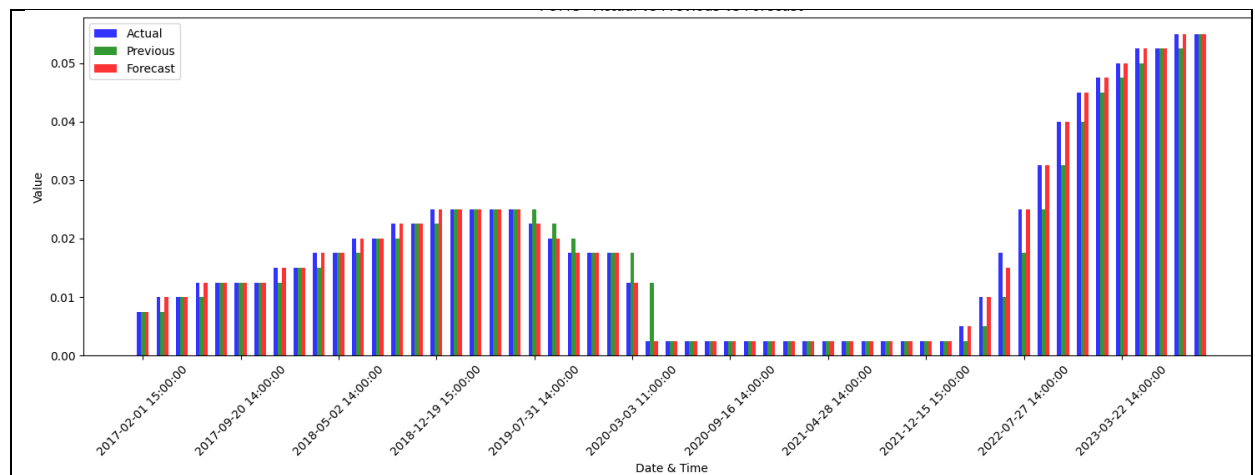

*Figure 5. Bar Plot of Actual, Previous, and Forecast FOMC interest rate*

*Note: This figure illustrates a bar graph comparing the actual, previous, and projected FOMC interest rates from February 2017 to September 2023. During this period, the FOMC convened for 55 meetings. Of these, the interest rate was raised on 18 occasions, reduced 5 times, and remained unchanged in 32 instances.*

Based on Tsay recommendation I used VAR model to analyze the effect of FOMC interest rate change on Bitcoin return. While OLS might only model the immediate impact of interest rate changes on Bitcoin returns, VAR contains lagged effects, allowing for a better understanding of how Bitcoin reacts over time to interest rate changes. The estimated VAR model is as follows:

$$\widehat{Bitcoin\_return}_t = 0.0506 - 1.968\,\widehat{FOMC_{Interest_{rate}}}_{t-1} + 0.189\,\widehat{Bitcoin\_return}_{t-1}$$

$$(0.02) \qquad (0.08) \qquad\qquad\qquad (0.016)$$

Based on the estimation result, FOMC interest rate change at time t-1 has negative significant effect on Bitcoin return. Therefore, we could conclude that FOMC interest rate change has negative effect on Bitcoin return. The result of impulse response of Bitcoin return to FOMC innovation, which is depicted in Figure 6 confirm these results too. Additionally, the effect of one period of Bitcoin return on current Bitcoin return is positive.

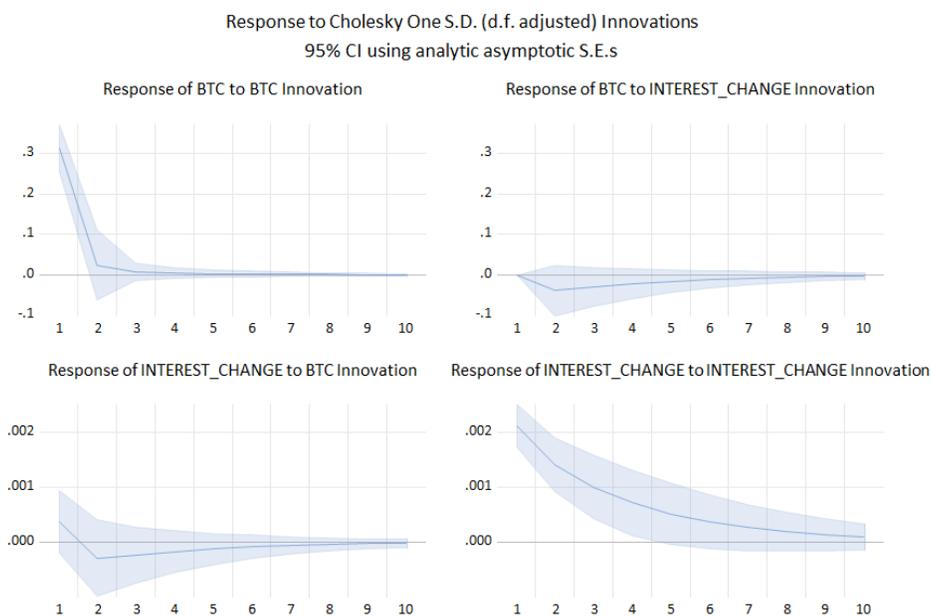

*Figure 6. Impulse Response of Bitcoin Return to FOMC Interest Rate Innovation*

*Note: This figure represents the impulse response of BTC to FOMC interest rate change innovation. As it is clear, BTC negatively response to innovation of FOMC interest rate change in the first two periods.*

# 6. CONCLUSION

The paper aimed to bridge a noteworthy gap in the economic literature by examining the influence of Monetary Policy Uncertainty (MPU) on cryptocurrency market dynamics, focusing on Bitcoin, using a sophisticated econometric approach. Using the Markov Switching Vector Autoregressive (MS-VAR) Model, our study discovered that Bitcoin's response to MPU varies according to different market states or regimes. Two distinct regimes were identified: a high volatility and a low volatility regime. In both these regimes, the results consistently indicated that increasing MPU leads to a decline in Bitcoin returns. These findings enrich our understanding of the intricate relationship between monetary policy and cryptocurrencies, and suggest that, much like traditional assets, the decentralized digital currency Bitcoin is not immune to the broader uncertainties in the global financial environment.

In addition to the findings, the paper addressed potential pitfalls and statistical concerns, such as residuals autocorrelation, endogeneity, and heteroskedasticity, ensuring that our results are robust and free from econometric biases. Specifically, the challenges of heteroskedasticity, which can often distort inference in time series analyses, were addressed and found to be of minimal concern, bolstering the reliability of our conclusions.

In closing, this research offers profound insights for policymakers, investors, and scholars. While the cryptocurrency market, represented here by Bitcoin, operates in a decentralized digital realm, its ties to traditional monetary policy dynamics are unmistakable. As digital assets continue to gain prominence, understanding these relationships will become increasingly crucial for both navigating investment decisions and shaping effective monetary policies.